# Rapid measurement of the local pressure amplitude in microchannel acoustophoresis using motile cells

Minji Kim[1], Rune Barnkob[2], and J. Mark Meacham[1a]

[1] *Department of Mechanical Engineering and Materials Science, Washington University in St. Louis, St. Louis, Missouri 63112, USA*
[2] *Heinz-Nixdorf-Chair of Biomedical Electronics, Department of Electrical and Computer Engineering, Technical University of Munich, TranslaTUM, 81675 Munich, Germany*

Acoustic microfluidics (or acoustofluidics) provides a non-contact and label-free means to manipulate and interrogate bioparticles. Owing to their biocompatibility and precision, acoustofluidic approaches have enabled innovations in various areas of biomedical research. Future breakthroughs will rely on translation of these techniques from academic labs to clinical and industrial settings. Here, accurate characterization and standardization of device performance is crucial. Versatile, rapid, and widely accessible performance quantification is needed. We propose a field quantification method using motile *Chlamydomonas reinhardtii* algae cells. We previously reported *qualitative* mapping of acoustic fields using living microswimmers as active probes. In the present study, we extend our approach to achieve the challenging *quantitative in situ* measurement of the acoustic energy density. *C. reinhardtii* cells continuously swim in an imposed force field and dynamically redistribute as the field changes. This behavior allows accurate and complete, real-time performance monitoring, which can be easily applied and adopted within the acoustofluidics and broader microfluidics research communities. Additionally, the approach relies only on standard bright-field microscopy to assess the field under numerous conditions within minutes. We benchmark the method against conventional passive-particle tracking, achieving agreement within 1 % for field strengths from 0 to 100 J m$^{-3}$ (0 to ~1 MPa).

## I. INTRODUCTION

Acoustofluidic devices are effective tools for manipulation of living cells in numerous biological and biomedical applications. However, consistency and reproducibility remain as barriers to their wider adoption due to the sensitivity of system performance on temperature, geometric/assembly tolerances, and variability in material properties.[1] Consequently, the translation of acoustic microfluidics from the research laboratory to clinical and industrial settings requires a robust means of experimental performance assessment and device calibration. Current experimental methods rely on laborious tracing of individual microparticle trajectories or complex characterization setups.[1-9] While computational models can guide device design and optimization, failure to capture real-world nonidealities limits their predictive power.[10-12] A rapid, accurate, and easy to implement quantitative measurement technique is needed.

Microscale objects placed in an acoustic field can be manipulated through two second-order acoustic effects, the acoustic radiation force due to wave-particle scattering and acoustic streaming due to the viscous attenuation of the wave, which acts on suspended particles via the Stokes drag force. Both effects scale linearly with the acoustic energy density $E^{\mathrm{ac}}$ (or squared pressure amplitude $p_{\mathrm{a}}^2$). Thus, the acoustic energy density and pressure amplitude are important metrics of device performance. Further, once the acoustic energy density is determined for a given microchannel, unknown acoustophysical properties (e.g., mass density, compressibility, and the acoustic contrast factor) for cells of interest can be determined and applied to manipulation/fractionation of target cells based on variations of these properties.[13, 14]

---

[a] *meachamjm@wustl.edu*

Considering objects larger than a certain size (typically 1–2 μm), the acoustic radiation force dominates particle motion.[4] Such motion is often termed acoustophoresis and is determined by balancing the acoustic radiation force and viscous drag from the fluid. Conventionally, theoretical expressions for the acoustophoretic motion of passive tracer particles are fit to experimentally observed particle trajectories to determine the acoustic energy density as a function of drive frequency and voltage;[2] however, tracing of individual particles is time-consuming and tedious. The passive particles rapidly achieve terminal distributions at low-pressure nodal locations corresponding to a given operating condition. Once they are moved, they cannot redistribute as conditions are varied, preventing continuous performance monitoring. For parametric studies, the entire channel must be flushed and reloaded with a fresh passive particle suspension prior to the next experimental condition. For each operating parameter to be tested (frequency, voltage, temperature, etc.), data collection can take hours to obtain a statistically significant number of trajectories. In addition, because the accuracy of the approach relies on knowledge of the Stokes drag force, the analysis must account for wall-corrections to the drag, requiring determination of the full three-dimensional (3D) particle trajectories.[5, 15]

We have previously demonstrated that active probes can address the limitations of passive particles, proposing the motile algae cell *Chlamydomonas reinhardtii* for this purpose.[16] Here, we extend our earlier work to establish a broadly applicable framework for use of *C. reinhardtii* cells to quantify performance of acoustofluidic devices (Fig. 1). *C. reinhardtii* cells swim within an imposed acoustic field so that the spatial distribution density of the swimming cells $\Lambda$ can be correlated to the field shape and strength (and thus, performance metrics including $E^{ac}$, acoustic potential $U$, and $p_a$) throughout the fluidic domain of a device. The dynamically responsive cells overcome limitations of conventional assessment methods, reaching a steady-state distribution during actuation but reverting to a uniform distribution in the absence of an external force field. To establish the utility of *C. reinhardtii* cells as a tool for performance assessment, we first accurately measure the acoustophysical properties of the cells. Mass density and compressibility determine an acoustic contrast factor that is combined with previously reported values for *C. reinhardtii* swimming capability (characterized by an intrinsic swimming velocity $U_c$ and reorientation time $\tau_c$) to allow correlation of the evolving cell distribution density to the acoustic field parameters. To demonstrate the method, the acoustic energy density and pressure amplitude within a simple straight microchannel driven at the first half-wavelength resonance are calculated as functions of drive voltage using the balance of swimming and acoustic radiation forces (see Fig. 1C). The resulting relationship matches that determined using passive tracer particles as a reference standard to within 1 %.

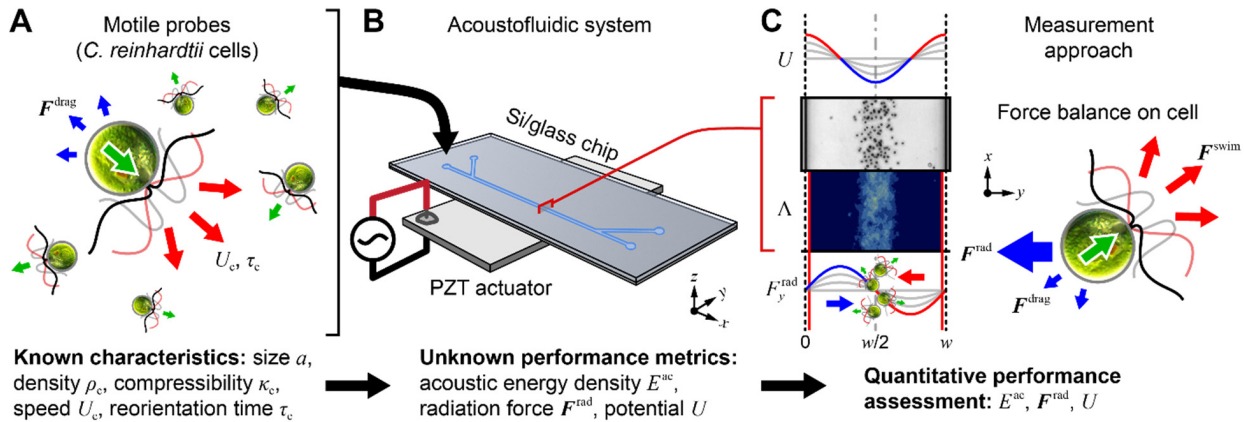

FIG. 1. (Color online). Concept for acoustic field quantitative performance assessment using motile cells as measurement probes. (A) *C. reinhardtii* cells with known swimming characteristics and acoustophysical properties (size, mass density, and compressibility). (B) Typical acoustofluidic system with unknown performance metrics. (C) The distribution density $\Lambda$ of swimming cells is used to measure performance of the acoustofluidic device by relating the intrinsic swimming capability of the cells to the acoustic radiation force.

## II. FIELD QUANTIFICATION METHOD
### A. CONCEPT

Particles suspended in an ultrasonic standing wave experience an acoustic radiation force due to scattering of the wave from the particles. For passive particles, the force leads to transient focusing to potential minima of the field (e.g., causing particle agglomeration at pressure nodal planes of a straight channel excited at a lateral resonance for positive acoustic contrast factor particles). In the ideal case, transient focusing is opposed only by viscous drag. Thus, if the acoustophysical properties (mass density, compressibility, and diameter) of the particles are known, particle motion can be correlated to acoustic field characteristics as introduced above. Conventional acoustic microfluidic device assessments have exploited this behavior to quantify the figures of merit for such devices (e.g., the acoustic energy density). The nearly perfect uniformity in size and properties of widely available polymeric microspheres is critical to the measurement accuracy; however, parametric studies involving passive particles are tedious.

Active particles have internal motors that generate an additional propulsive force to counteract the acoustic radiation force (see Fig. 1; *C. reinhardtii* cell as a representative active particle). In the absence of an external force field, particle motion is characterized by a swimming speed $U_c$ and reorientation time $\tau_c$, which dictate a characteristic run length before turning. When placed in a standing wave, individual particles will continue to swim relative to the radiation force (assuming it does not interfere with the propulsive mechanism), with the external field either assisting or impeding the directed motion of the particle (Fig. 1C). Active particles will continuously probe the potential minima of the field, over time fully exploring regions bounded by a threshold force (i.e., the radiation force equal and opposite in direction to the propulsive force). Thus, a population of active particles can map the contours of a field delineated by this force balance. If the field strength is increased/decreased, particles become more/less confined within the potential minima, yielding additional field contours. Again, if the size, acoustophysical properties, and swimming characteristics are known, the steady-state distribution density of particles can be related to the field shape and strength (e.g., to quantify acoustic energy density); however, active particles are suitable for dynamic measurements, reflecting changes in the field in real-time and allowing parametric performance characterization without reloading.

### B. Active particle of choice: the *C. reinhardtii* cell

Many biological cells would make poor measurement probes due to their inherent heterogeneity. In contrast, the unicellular alga *C. reinhardtii* is an excellent candidate active particle for demonstration of our method. Vegetative *C. reinhardtii* cells are driven to differentiate into haploid gametes of a single-mating type by suspension in nitrogen-depleted medium. The cell life cycle is halted, creating a synchronized population of uniformly sized (diameter = 8.0 ± 1.1 μm, *n* = 138) cells that are well-suited to acoustophoretic manipulation at low-MHz frequencies. The oscillatory motion of propulsive cilia is relatively unaffected due to its nanoscale cross section (~250-nm diameter). We have previously reported a swimming speed of 92 ± 5 μm s$^{-1}$ and reorientation time of ~0.75 s for wild-type *C. reinhardtii* cells (strain CC-125).[17] The resultant propulsive force is ~10 pN, which is consistent with earlier measurements.[18] The cells can traverse microfluidic domains in a few of seconds, generating a comparable force to typical acoustic radiation forces on cell-sized objects.[19] To realize the full potential of the method, the present work includes accurate measurement of *C. reinhardtii* cell acoustophysical properties to allow determination of the acoustic contrast factor (mass density $\rho_c$ = 1119 ± 11 kg m$^{-3}$ and compressibility $\kappa_c$ = 386 ± 3 TPa$^{-1}$; see Materials and Methods section). From a practical perspective, laboratory culture and genetic modification of *C. reinhardtii* cells are straightforward due to their role as model organisms in the study of ciliary diseases in humans. As algae cells, they can be prepared and maintained on a standard laboratory bench. Thus, they are accessible even to researchers unfamiliar with biological techniques.

### C. Particle manipulation in an acoustic wave field

Acoustofluidic microsystems are typically actuated harmonically, either through surface acoustic waves travelling on a piezoelectric substrate or through a bulk piezoelectric transducer. Time-harmonic oscillating pressure and velocity fields [$p_1(\boldsymbol{r})$e$^{-i\omega t}$ and $\boldsymbol{v}_1(\boldsymbol{r})$e$^{-i\omega t}$] are established when the actuator is driven harmonically ($\omega = 2\pi f$ is the corresponding angular frequency). These first-order acoustic fields drive two second-order phenomena, the acoustic radiation force and acoustic streaming. The acoustic radiation force arises due to scattering of the acoustic waves as they encounter suspended particles that are moved to acoustic potential minima as a result. Acoustic streaming is driven by viscous attenuation of the acoustic waves, which results in a steady bulk fluid flow generating a drag force on the particles. As the fluid drag force and radiation force scale differently with particle size, the motion of small particles (~< 3 μm) is dominated by the acoustic streaming drag, while the motion of relatively larger particles (~> 3 μm) is dominated by the acoustic radiation force.[4] Since the size (~ 8 μm) of *C. reinhardtii* cells exceeds the transition threshold for which the acoustic streaming becomes negligible (~ 3 μm), the radiation force $\boldsymbol{F}^{\text{rad}} \equiv \boldsymbol{F}^{\text{rad}}(\boldsymbol{r})$ is the dominant acoustic effect, and the motion of a cell with swimming force $\boldsymbol{F}^{\text{swim}} \equiv \boldsymbol{F}^{\text{swim}}(t)$ can be described as

$$\frac{4\pi}{3}a^3\rho_c\partial_t\boldsymbol{u} = \boldsymbol{F}^{\text{rad}} + \boldsymbol{F}^{\text{drag}} + \boldsymbol{F}^{\text{swim}}, \tag{1}$$

where $a$ is the particle radius, $\rho_c$ is the cell mass density, $\boldsymbol{u} \equiv \boldsymbol{u}(\boldsymbol{r}, t)$ the cell velocity, $\boldsymbol{F}^{\text{drag}} = -\zeta\boldsymbol{u}$ is the viscous drag force, and $\zeta$ is the hydrodynamic drag coefficient. $\boldsymbol{F}^{\text{swim}} = -\zeta\boldsymbol{u}^{\text{swim}}$ is the self-propulsive swim force of a cell, where $\boldsymbol{u}^{\text{swim}} = U_c\boldsymbol{q}$ is the intrinsic swim velocity of an isolated swimming cell and where $\boldsymbol{q} \equiv \boldsymbol{q}(t)$ is the swim direction. Neglecting wall effects and considering the cell to be nearly spherical, the hydrodynamic drag coefficient can be described as $\zeta = 6\pi\eta a$ with $\eta$ being the dynamic viscosity of the suspension medium. Furthermore, as the characteristic time of acceleration $\tau \sim \rho_c a^2\eta$ is small (~ 10 μs) in comparison to the experimental time scale characterizing the motion of the cells (> 1 ms), we neglect inertial effects, and the cell motion can be described as

$$\boldsymbol{u} = \frac{\boldsymbol{F}^{\text{rad}}}{6\pi\eta a} + \boldsymbol{u}^{\text{swim}}. \tag{2}$$

The acoustic radiation force for a spherical particle of compressibility $\kappa_c$ and size much smaller than the acoustic wavelength ($a \ll \lambda$) is given by Karlsen and Bruus,[20]

$$\boldsymbol{F}^{\text{rad}} = -\pi a^3\left\{\frac{2\kappa_o}{3}\text{Re}[f_0^*p_1^*\boldsymbol{\nabla}p_1] - \rho_o\text{Re}[f_1^*\boldsymbol{v}_1^* \cdot \boldsymbol{\nabla}\boldsymbol{v}_1]\right\}, \tag{3}$$

where $p_1$ and $\boldsymbol{v}_1$ are the first-order pressure and velocity, respectively, and where $\kappa_o$ is the fluid compressibility, $\rho_o$ is the fluid mass density, the asterisk denotes complex conjugation, Re[$A$] denotes the real part of $A$, and where the two dimensionless scattering coefficients $f_0$ and $f_1$ are given by

$$f_0 = 1 - \frac{\kappa_c}{\kappa_o} \text{ and } f_1 = \frac{2(\rho_c - \rho_o)}{2\rho_c + \rho_o}. \tag{4}$$

Here, we neglect thermoviscous corrections to the scattering coefficients as these corrections are small (relative to the error in measured compressibility and mass density) for the cells and suspension medium (see above and Materials and Methods section).

For hard-walled microchannels, the channel sidewalls support a standing pressure wave, and the radiation force reduces to[21-25]

$$\boldsymbol{F}^{\text{rad}} = -\frac{4\pi}{3}a^3 \nabla \left[ f_0 \frac{1}{2} \kappa_o \langle p_1^2 \rangle - f_1 \frac{3}{4} \rho_o \langle v_1^2 \rangle \right]. \tag{5}$$

The acoustic field generated within the fluidic channel is dependent on the channel geometry. In the present study, we use a straight channel with length $l$, width $w$, and depth $d$ in the $x$-, $y$-, and $z$-coordinate directions, respectively. For a channel with $l \gg w > d$ actuated at the lateral ($y$-coordinate direction) first half-wavelength resonance, a single potential minimum is formed at the channel midline (i.e., at $y = w/2$). In this one-dimensional (1D) case with $p_1(y) = p_a \cos(k_y y)$ and pressure amplitude $p_a$, the acoustic radiation force further simplifies to[21-25]

$$F_y^{\text{rad}} = 4\pi \Phi a^3 k_y E^{\text{ac}} \sin(2k_y y), \tag{6}$$

where $\Phi = f_0/3 + f_1/2$ is the acoustic contrast factor and $E^{\text{ac}} = p_a^2/(4\rho_o c_o^2)$ is the acoustic energy density. Note that due to the second-order nature of the force, it is period doubled, and a half-wavelength standing pressure wave results in a full sine-wave force field.

### D. Swimming cell confinement in an ultrasonic standing wave

We applied Eqs. (2) and (6) to perform molecular dynamics simulations of $10^5$ 'swimming' point particles (cells) under acoustic actuation in a straight microchannel with a simulation domain sized to the field of view used in experiments ($l = 834$ μm, $w = 375$ μm; see Materials and Methods section). The model cells reoriented randomly every second or upon encountering a channel wall. Wall conditions were imposed at the boundaries in the $y$- and $z$-**coordinate** directions, while cells were subject to periodic boundary conditions in the $x$-**coordinate** direction along the channel length. Cell positions were recorded after 20 s of swimming in the acoustic field. Spatially averaged model cell distribution densities for three channel acoustic energy densities are compared to experimentally observed cell distributions to provide insight into how acoustic confinement affects the spread of a swimming cell population (see Fig. 2).

When the acoustic radiation force is smaller than the swim force ($|F_y^{\text{rad}}| < |F_y^{\text{swim}}|$), the propulsive force of the cells overcomes the radiation force, and no acoustic trap is formed (Subthreshold trap distribution, see Fig. 2A). Under these conditions, free swimming cells explore the entire width of the channel. When the maximum value of the radiation force is equal to the swim force ($|F_y^{\text{rad}}|_{\max} = |F_y^{\text{swim}}|$), cells are no longer able to overcome the acoustic radiation force and begin to amass near the midline of the channel (Bimodal trap distribution, see Fig. 2B). Due to the sinusoidal shape of the acoustic radiation force across the channel width, maximum values of $|F_y^{\text{rad}}|$ occur at $w/4$ and $3w/4$. Thus, the largest trap width that can be formed using *C. reinhardtii* cells at the first half-wavelength resonance of a straight channel is half the channel width $w/2$. At higher voltages, $F_y^{\text{rad}}$ exceeds $F_y^{\text{swim}}$ throughout most of the channel, and cells achieve a suprasaturated trap distribution where further confinement is not possible (see Fig. 2C).

Trap boundaries form at locations where the swim force and the acoustic radiation force balance each other, and consequently, the trap width decreases with increasing acoustic energy density (applied acoustic drive voltage). Interestingly, cells appear to 'stack up' at the these boundaries suggesting that the trap is not large compared to the run length of the cell (i.e., it is a 'strong' trap).[26] Since $|F_y^{\text{rad}}| = |F_y^{\text{swim}}|$ ($u_{c,y} = 0$) at the trap boundaries, the 1D versions of Eq. (2) and Eq. (6) can be used to express the acoustic energy density in terms of the $y$-location $y_{\text{bd}}$ of the acoustic trap boundary,

$$E^{\text{ac}} = \zeta U_o \left[4\pi \, \Phi \, a^3 k_y \, |\sin(2k_y y_{\text{bd}})|\right]^{-1}. \quad (7)$$

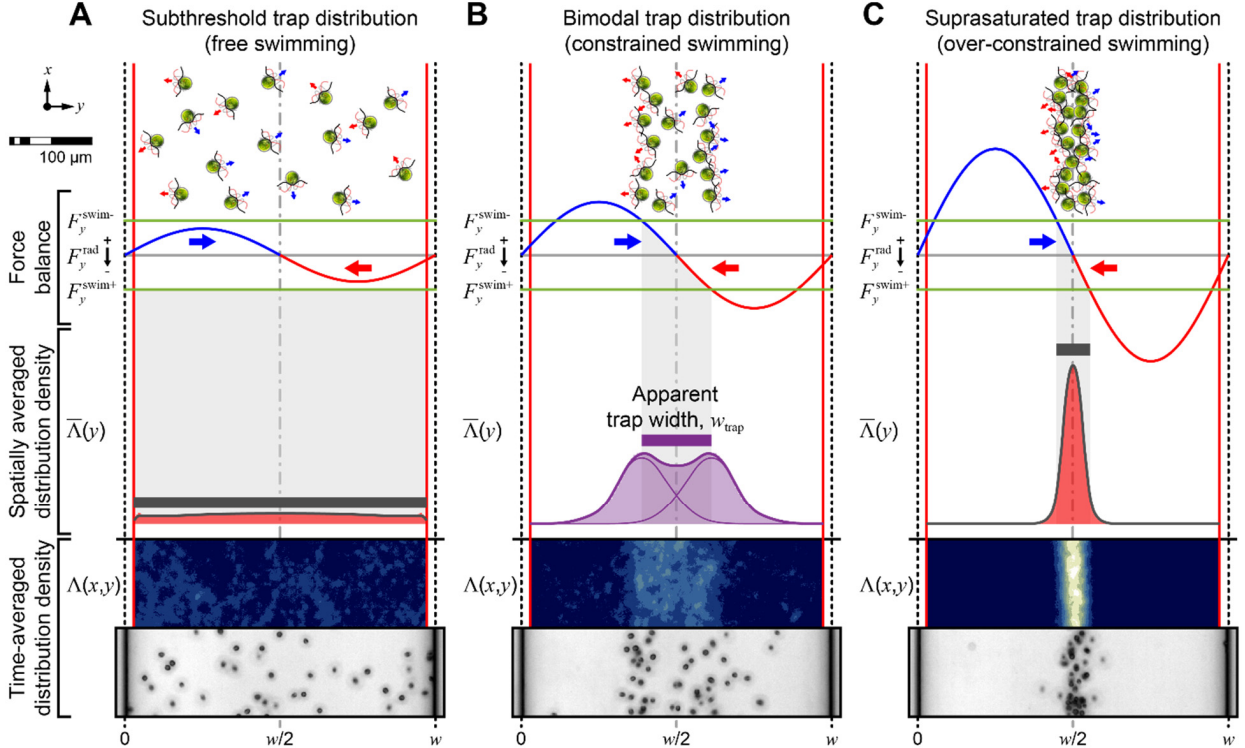

FIG. 2. (Color online). Concept illustration and molecular dynamics simulations of *C. reinhardtii* cell confinement in acoustic fields of three different strengths. (A) $|F_y^{\text{rad}}| < |F_y^{\text{swim}}|$ and *C. reinhardtii* cells overcome the acoustic radiation force to explore the entire microfluidic channel. (B) The field strength exceeds the threshold ($|F_y^{\text{rad}}|_{\max} = |F_y^{\text{swim}}|$) where *C. reinhardtii* cells begin to amass near the channel midline. The cells exhibit a bimodal distribution due to cell 'stacking' at the trap boundaries. (C) The acoustic radiation force amplitude is significantly larger than the swim force ($|F_y^{\text{rad}}|_{\max} > |F_y^{\text{swim}}|$), and the acoustic trap width decreases until reaching a suprasaturated trap density where no further confinement is possible.

Thus, we can determine the acoustic energy density *in situ* by experimentally measuring the acoustic trap boundaries (i.e., the apparent trap width $w_{\text{trap}}$; see Fig. 2B) using cells of known acoustophysical properties and swimming characteristics. Note that the above analytical predictions are based on simple point-particles representing isolated swimming cells, and thus, the model neglects experimental complexities such as cell-cell and cell-wall interactions, discrete image pixel recordings, and the effects of cells blocking the observation of one another.

### E. Implementation

Having selected an active probe with known acoustophysical properties, the user can characterize performance of an acoustofluidic device. *C. reinhardtii* cells are simple to culture and maintain (even in nonbiological labs). Cells are taken from culture on the bench, suspended in a nitrogen-depleted medium, and ready for experimental use (e.g., cells have become mature gametes) in a few hours. Once these active probes are loaded, numerous operating parameters can be tested in series without changing the sample, significantly reducing data collection time. The user may 'sweep' any operating parameter to investigate how the acoustic field evolves. In the present work, the voltage input is gradually increased, and the line shape representing the cell distribution density is related to the acoustic energy density within the field of view of a straight microchannel actuated at its first half-wavelength resonance (Fig. 2). Once the local acoustic energy density is determined, other device performance metrics (e.g., pressure amplitude) can be derived.

### III. PROOF OF CONCEPT AND VALIDATION
#### A. Materials and Methods
##### 1. Microfluidic chip and experimental set-up

Conventional microfabrication processes were used to fabricate a silicon-glass chip with a rectangular fluidic channel ($l$ = 30 mm, $w$ = 0.375 mm, $d$ = 50 μm). The channel was etched in a 300-μm-thick silicon wafer following a two-step deep reactive ion etch (DRIE; front side channel, back-side inlet/outlet through). A 500 μm thick Borofloat® 33 cover glass was anodically bonded to enclose the channel. A PZT-8 piezoelectric transducer (14 mm × 24 mm × 0.75 mm; APC 880, American Piezo Ceramics) was used to excite vibrational and acoustic waves in the device. The transducer was bonded to the chip with a thin layer of epoxy. Female luer ports (Fluidic 631, ChipShop) were epoxied to the inlet and outlet holes to facilitate reversible tubing connections. The chip was mounted on a temperature-controlled stage insert (PE100, Linkam). To ensure consistent and repeatable performance, a custom-designed 3D printed chip holder (Prusament PLA, Prusa Research) was used to lock the chip in place on the stage insert. An automated fluid delivery system comprising a pressure control unit (MFCS-EZ, Fluigent), a multiport rotary valve (M-Switch, Fluigent), and a flow sensor (Flow Unit, Fluigent) was used to load particles and cells into the chip.

##### 2. Cell culture and sample preparation

*C. reinhardtii* cells were provided by the Dutcher Lab at Washington University in St. Louis. Wild-type (strain CC-125) cells were incubated on agar plates at 25°C under constant lighting for 48 hours, as previously reported.[27] To provoke cilia growth and to obtain a population of gametes (rather than vegetative cells) with uniform size and motility, cells were suspended in a nitrogen-depleted liquid growth medium for three hours before an experiment. During this time, the cell sample was kept on a cell rocker (Rotator Genie, Scientific Industries) at a low speed to prevent clumping. Finally, the cell sample was centrifuged at 1000 g for 5 min, and excess medium was removed to attain a sample concentration of ~10 × 10$^6$ cells per mL.

##### 3. Measurements of acoustophysical properties

Still images of cells were analyzed using imaging software (ZEN software, Zeiss) to obtain a representative cell size distribution. The viscosity of the medium at 20°C was measured using a rheometer (AR-G2, TA Instruments) with the shear rate of 1/120 s$^{-1}$. The mass density and speed of sound of the growth medium at 20°C were evaluated using a density and sound velocity meter (DSA 5000 M, Anton Paar). The compressibility of the medium was inferred from the mass density and speed of the sound measurements.

For mass density measurements of the cells, *C. reinhardtii* cells were first immobilized by removing the cilia using an acid shock.[28] The nanoscale cilia (~250 nm in diameter, 10 μm in length) have a negligible effect on the cell mass density. A cell solution of 1 mL was treated with acetic acid (7 μL; 0.5 N) and vortexed for 45 s. After buffering with potassium hydroxide (3.5 μl; 0.5 N), cells were centrifuged at 1000 g for 5 min. Finally, deciliated cells were resuspended in fresh medium. Since cells start regrowing cilia after the medium exchange, measurements were taken immediately after the deciliation process. Density gradient centrifugation was used to obtain the mass density. The speed of sound in a neutrally buoyant cell suspension at 20°C was measured at various volume fractions to obtain the average compressibility of *C. reinhardtii* cells as described by Cushing *et al.*[29] The mass density measurements were used to prepare the neutrally buoyant solutions by adding an appropriate amount of iodixanol solution (OptiPrep, Axis-Shield Poc AS) to the cell sample. Following the Wood's equation, the relationship between the cell volume fraction and the speed of sound was used to deduce the average compressibility of the cells.

### 4. Observation of polystyrene bead trajectories in acoustic fields

Acoustic focusing of 5-μm polystyrene (PS) beads (Phosphorex) was observed using an inverted microscope (Axio Observer z.1, Zeiss) and a 3-Megapixel camera (Axiocam 503, Zeiss). The protocol reported by Barnkob *et al.*[2] was followed. PS beads were suspended in the algae growth medium to ensure the same fluid properties as those of the cell trapping experiments. The bead solution was loaded into the chip using the automatic fluid delivery system. The ultrasound field was turned on at the first half-wavelength resonance of the microchannel, $f$ = 1.811 MHz (33522A, Agilent; 2100L, ENI). Migration of the beads to the acoustic potential minimum at the channel midline was recorded at 38 fps. The channel was flushed and loaded with new beads, and focusing experiments were repeated until a statistically significant number of bead trajectories was collected at each voltage step (>100). This procedure was performed for five actuation voltages: 1.68 $V_{pp}$, 2.53 $V_{pp}$, 3.45 $V_{pp}$, 4.25 $V_{pp}$, and 5.06 $V_{pp}$. Output voltage was monitored using a PC oscilloscope (PicoScope 2204A; Pico Technology). The temperature was maintained at 20°C. The transversal trajectories of the beads were analyzed using the General Defocusing Particle Tracking (GDPT) method using the open-source *DefocusTracker* implementation, see more via https://defocustracking.com/.[30, 31]

### 5. Observation of C. reinhardtii cell distributions in acoustic fields

*C. reinhardtii* cells were loaded into the microfluidic chip, and static trapping experiments were performed using the same setup used for the transient PS bead focusing. The field of view used in experiments was $l \times w$ = 834 μm × 375 μm. Again, the device was actuated at the first half-wavelength resonant frequency $f$ = 1.811 MHz. Since swimming cells revert to a uniform distribution in between each voltage, there was no need to flush the channel as was done in between PS bead experimental conditions. Once the sample of cells was loaded, the entire voltage range was actuated sequentially, i.e., the voltage was increased from 0 $V_{pp}$ to 8.2 $V_{pp}$, at ~0.40 $V_{pp}$ increments. Each voltage step was 'on' for 10 s, and acoustic actuation was turned off for 5 s in between each voltage step to allow cells to redistribute across the channel, ensuring a uniform distribution was achieved at the start of the subsequent step. The evolving cell distributions were recorded at 5 fps. Drive voltage was automatically controlled using a custom Python script that also logged the start time for each voltage step. Actuation and image acquisition time points were correlated to assign a voltage to each image in a sequence. The temperature was maintained at 20°C, which is in the range of the ideal conditions for *C. reinhardtii* cells. After each full voltage sweep, the chip was flushed and reloaded to obtain experimental replicates. The procedure was repeated a total of 5 times over 1 h.

### 6. Cell distribution image processing and analysis

The first twenty images of each voltage sequence (i.e., before the first voltage step was actuated) were used to create a background image. Captured grayscale brightfield images were inverted, and the inverted background image was subtracted from subsequent images in the series. Initially, areas occupied by cells were dark gray or black against a light background. After inversion, those areas were light gray or white with high pixel counts (closer to 16,383 on the 0-16,383 14-bit pixel value scale) against a dark background with low pixel counts (closer to 0). The pixel counts in each image were summed for images at the same voltage step (35 images). Consequently, regions where cells dwelled for the longest time had the highest summed pixel count. Finally, the pixel counts were normalized using the highest summed pixel count within an experimental trial, i.e., the highest normalized summed pixel count was 1. Distribution maps visualizing the cell distribution densities (light: high density, dark blue: low) were created using the *davos* sequential scientific color map.[32]

## B. RESULTS AND DISCUSSION

### 1. *Acoustophysical properties of C. reinhardtii cells and their growth medium*

To calculate the relevant forces (i.e., the acoustic radiation force $F_y^{\text{rad}}$, swimming force $F_y^{\text{swim}}$, and the drag), the size distribution of the cells, viscosity of the growth medium, and mass density and compressibility of both the cells and growth medium are needed. The cell growth medium exhibits water-like properties with a viscosity of 1.02 ± 0.05 mPa s, mass density 997 kg m$^{-3}$, and compressibility 448 TPa$^{-1}$, all measured at 20°C. The microscale *C. reinhardtii* cell body (radius $a$ = 4.0 ± 0.6 μm ($n$ = 138) is well-suited to generating a large primary acoustic radiation force. The mass density and compressibility of *C. reinhardtii* cells at 20°C are 1119 ± 11 kg m$^{-3}$ and 386 ± 3 TPa$^{-1}$, respectively. With these material properties, the acoustic contrast factor is $\Phi$ = 0.084 ± 0.004 ($f_0$ = 0.138 ± 0.007 and $f_1$ = 0.075 ± 0.006; see Eq. 4). For comparison, the mass density and compressibility of red blood cells are 1101 ± 13 kg m$^{-3}$ and 334 ± 2 TPa$^{-1}$ and those of breast cancer cells (MCF-7) are 1055 ± 1 kg m$^{-3}$ and 373 ± 1 TPa$^{-1}$.[29]

### 2. *Focusing behavior of C. reinhardtii cells in an ultrasonic standing wave*

The distribution of *C. reinhardtii* cells in an acoustic field reflects the field shape and strength in real time (Fig. 3Ai). A series of brightfield images taken at the same voltage step are averaged to create a distribution map that provides a qualitative description of the field (Fig. 3Aii). Before quantitative determination of the field characteristics, the time-averaged distribution map is segmented and aligned to best represent the assumed 1D ultrasonic standing wave. At the first half-wavelength resonance of the straight channel, the potential minimum forms at the channel midline where cells congregate under the acoustic actuation. In the ideal case, the acoustic potential is constant along the *x*-coordinate direction, varying only across the *y*-coordinate direction; however, a small variation along the *x* direction is observed experimentally. Images taken at the highest voltage, which exhibits a single peak at the channel midline, are divided into eight sections along the *x* direction. Locations of the segment-specific peaks are determined, and the eight segments are shifted left or right to align these peak locations. Time-averaged heat maps for the entire series of voltage steps are shifted accordingly (Fig. 3Aiii).

The swimming behavior of *C. reinhardtii* is described as a random walk.[33, 34] The wild-type *C. reinhardtii* cells (CC-125) swim ballistically (i.e., in an almost straight line) for ~1 s (reorientation time) prior to making a turn at a random angle.[17, 33] The distance travelled before turning is called the run length. In a weak trap, where the size of the trap is much larger than the microswimmer run length, the cells reorient before reaching the trap boundary. However, in a strong trap, where the trap size is comparable to or smaller than the run length, the swimmers encounter the trap boundary before they can reorient. When a swimming cell reaches a trap boundary (defined as the location where $|F_y^{\text{ac}}|$ = $|F_y^{\text{swim}}|$), the cell dwells at the boundary until the reorientation time period ends.[26] Thus, cells are more likely on average to be present in regions near the trap boundaries. The distribution density of these boundary populations is well-represented by a Gaussian function (see Fig. 3B). As discussed earlier, we have previously determined the characteristic velocity (~90 μm s$^{-1}$) and reorientation time time (~0.75 s) for the *C. reinhardtii* cells, yielding a run length of ~70 μm. Typical acoustofluidic traps in the 1–10 MHz range span from 375 μm to 37.5 μm in width ($+|F_y^{rad}|_{max}$ to $-|F_y^{rad}|_{max}$ or node to antinode distance). In the present study, we observe trap widths (delineated by boundary populations of cells) of between approximately 20 μm and 95 μm. Thus, the trap width and *C. reinhardtii* cell run length are of the same order, and the experimental conditions are well within the strong strap regime.

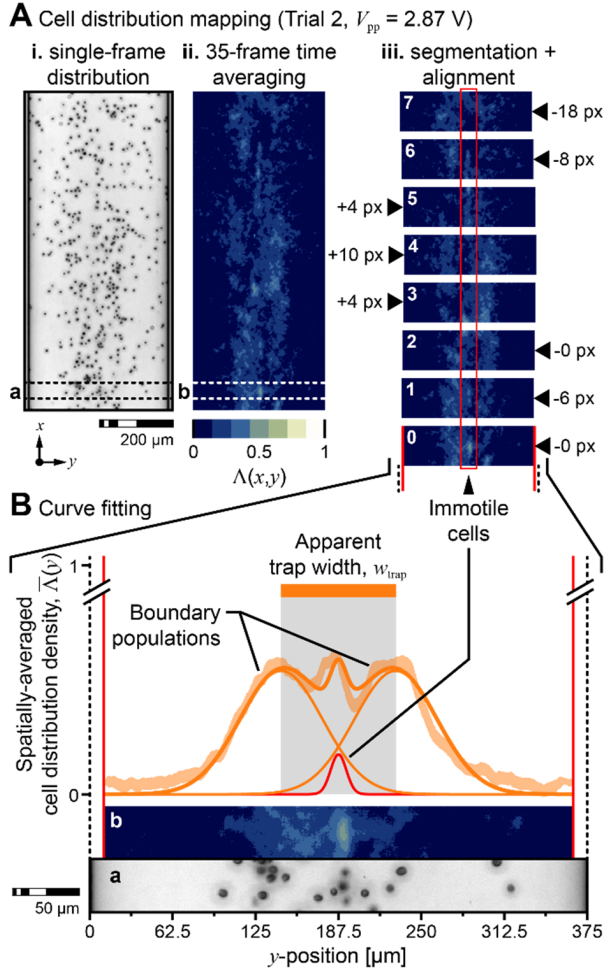

FIG. 3. (Color online). Image processing and curve fitting of cell distribution images. (A) *C. reinhardtii* cell distribution mapping (time averaging, segmentation, and alignment). (B) Curve fitting using a three-peak Gaussian fit. Primary left and right peaks represent cell stacking at the trap boundaries (defining an apparent trap width $w_{trap}$). A smaller central peak composed of immotile cells develops at the trap midline. Cell distribution (a) and heat map (b) slices are included for reference.

A small number of immotile cells (whether dead or dysfunctional) exist in each sample. Since these cells lack an intrinsic swim force, they cannot explore the extents of the larger trap and immediately focus to the trap midline under the action of an ultrasonic standing wave. Once driven to the midline, these cells do not redistribute as the voltage signal is varied. To account for this effect, we use a three-peak Gaussian fit to represent the spatially averaged (along the *x*-coordinate direction spanning the field of view) cell distribution density (see Fig. 3B). At the first voltage step beyond the threshold where cells begin to aggregate (~2.9 $V_{pp}$), the two boundary peaks and the middle peak formed by the immotile cells are identified by fitting the following equation,

$$Y = a_i \exp\left[-\left(\frac{x - b_i}{c_i}\right)^2\right] + a_{\text{mid}} \exp\left[-\left(\frac{x - b_{\text{mid}}}{c_{\text{mid}}}\right)^2\right] + a_i \exp\left[-\left(\frac{x + b_i}{c_i}\right)^2\right], \quad (8)$$

with the outer primary peaks set to be symmetric about the identified midline. For subsequent voltage steps, the values of $a_{\text{mid}}$, $b_{\text{mid}}$, and $c_{\text{mid}}$ are held constant (assuming no further focusing of the immotile cell population occurs after the first fit condition), while the fitting coefficients $a_i$, $b_i$, and $c_i$ were determined for each voltage step $i$. Finally, we define the trap width $w_{\text{trap}}$ as the peak-to-peak distance between the outer primary peaks.

### 3. Measurement of the acoustic energy density using C. reinhardtii cells

Using *C. reinhardtii* cells as active probes, the evolving acoustic field was observed as voltage was varied from 0 to 8.20 $V_{\text{pp}}$. The resulting distributions correspond to three distinct confinement regimes: subthreshold, curve fitting, and suprasaturation (see Figs. 2 and 4). In the subthreshold range, where the applied voltage is below the threshold for confinement, the swimming force of the cells is greater than the acoustic radiation force. Immotile cells with zero swim force are moved to the channel midline (the potential minimum of the half-wavelength standing wave), but swimming cells overcome the acoustic radiation force and no apparent acoustic trap is formed. Just beyond the confinement threshold at 2.87 $V_{\text{pp}}$, cells begin to aggregate within the region between the radiation force maxima that delineate the trap ($+|F_y^{\text{rad}}|_{\text{max}}$ to $-|F_y^{\text{rad}}|_{\text{max}}$). The three-peak Gaussian fitting approach described above is used to define the trap boundary locations and apparent trap width $w_{\text{trap}}$. As the voltage is gradually increased, the boundary locations shift closer to the midline, and the trap width narrows, reflecting the increase in field strength. When $w_{\text{trap}}$ becomes smaller than the effective cell size (body diameter plus cilia length, ~18 μm), cells are considered tightly packed and neighboring cells constrain further confinement. The $w_{\text{trap}}$ cannot decrease significantly beyond a minimum value corresponding to trap saturation at a voltage of ~6 $V_{\text{pp}}$ (see Fig. 4). It becomes difficult to achieve a three-peak fit, and even if three peaks can be resolved, any further decreases in $w_{\text{trap}}$ are not solely a function of the increasing trap strength. In fact, cells are so tightly packed in the suprasaturation range that a single peak fit is unavoidable. Because the apparent trap width is ill-defined in the subthreshold and suprasaturation ranges, these experimental conditions are excluded from the performance analysis, i.e., conditions yielding trap widths smaller than 18 μm are not used for $E^{\text{ac}}$ calculations (see Fig. 5).

Analysis of the $w_{\text{trap}}$ variation with increasing voltage suggests that there could be a potential change in swimming speed as experimental time progresses. Note that although the 5 experimental repeats were performed with different *C. reinhardtii* samples, samples were taken from the same stock suspension so that the last trial (T5 in Fig. 5) occurred almost 1 h after the first trial (T1, Fig. 5). The trap width as a function of applied voltage $V$ exhibits an exponential decay as shown in Fig. 5. However, the decay is more severe for the later trials. We suspect that the swimming capability of the cells may decrease slightly over time. In the current study, the free-swimming cells were recorded at a high enough frame rate for the swim speed calculation only at the beginning of the first experiment. This can result in an incorrect $\boldsymbol{F}_{\text{swim}}$ calculation. In the future, *in situ* swimming speed determination should be performed before each trial to ensure an accurate $\boldsymbol{F}_{\text{swim}}$ measurement. Though this requires an additional analysis step, each $\boldsymbol{F}_{\text{swim}}$ calculation can be completed in less than 1 min.

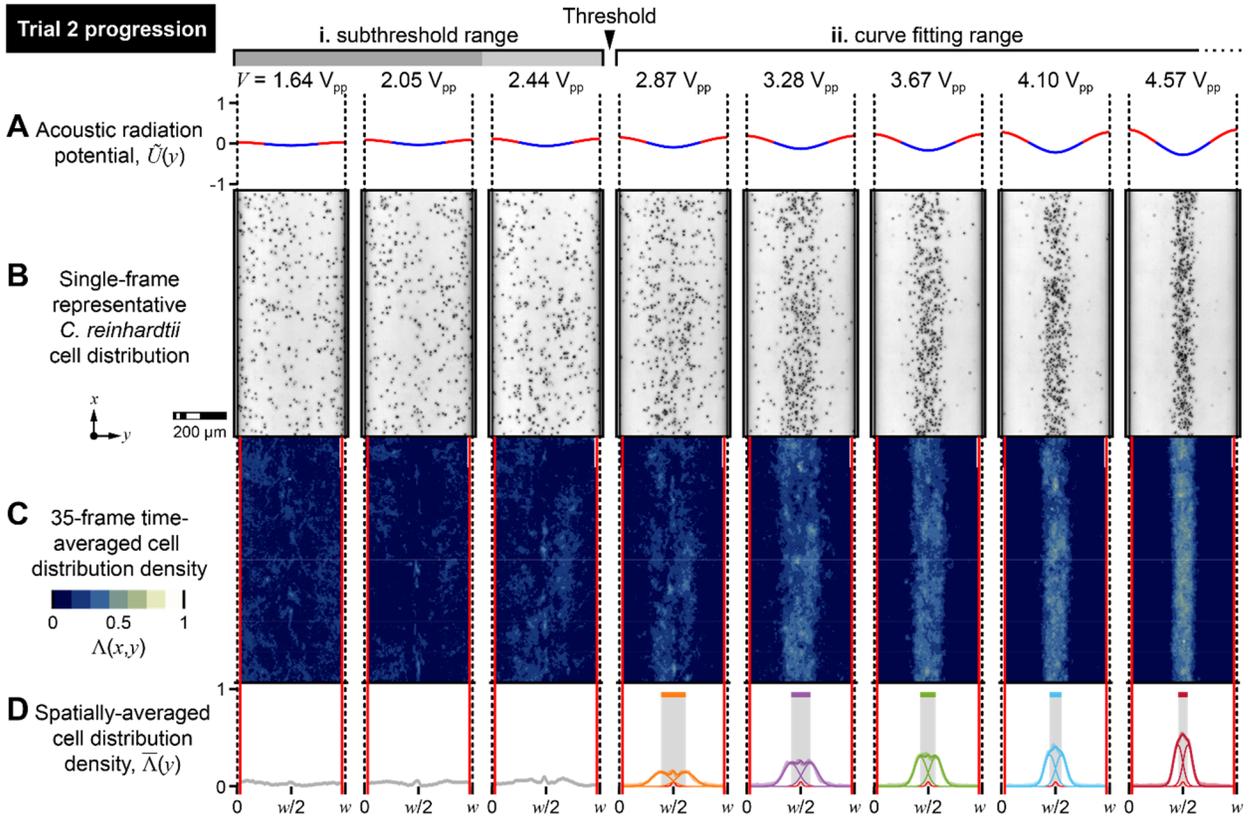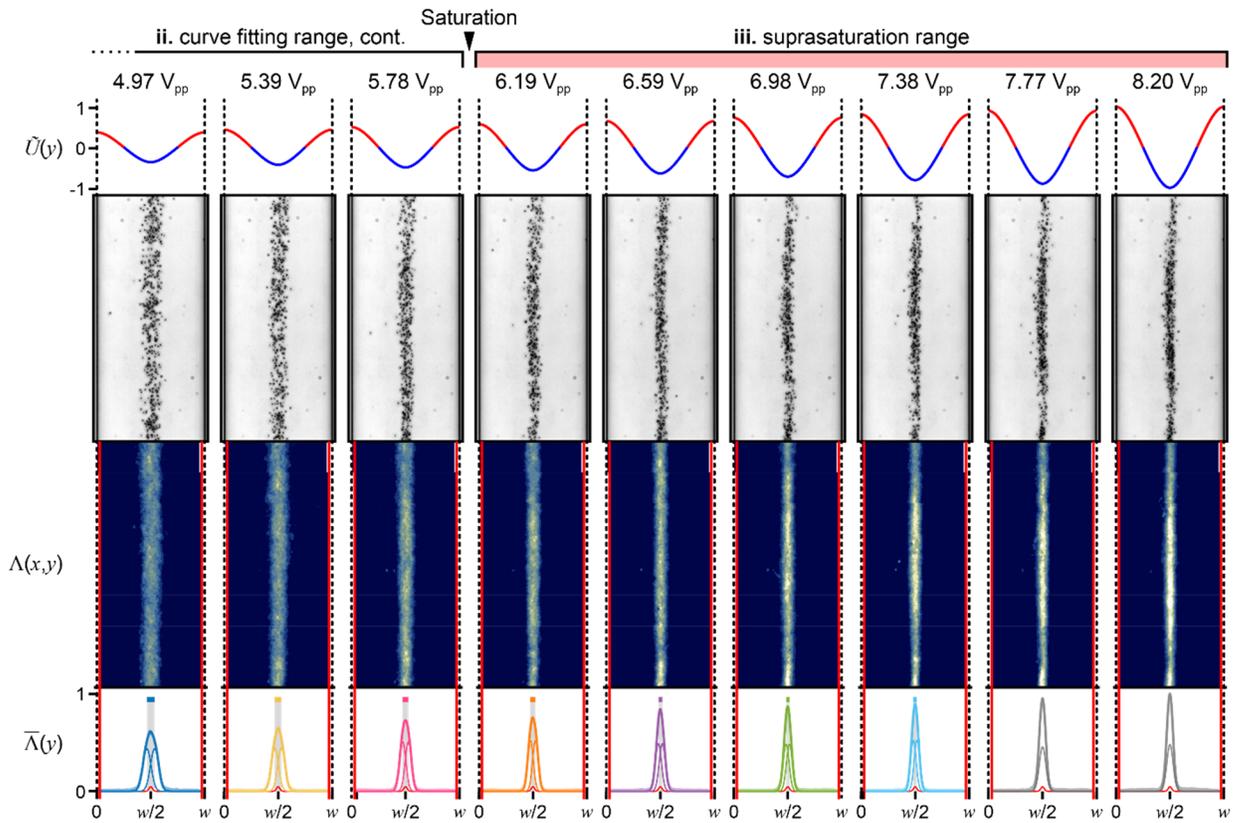

FIG. 4. (Color online). (A) Acoustic radiation potential $\tilde{U}$, (B) single-frame example of the *C. reinhardtii* cell distribution, (C) heat map representing a 35-frame time-averaged *C. reinhardtii* cell distribution density $\Lambda$ normalized to the highest intensity for the entire experimental trial, and (D) spatially averaged (along the *x*-coordinate direction spanning the field of view) *C. reinhardtii* cell distribution density $\bar{\Lambda}$ for each voltage step of the experimental Trial 2 (starting 21 min after sample preparation). The three confinement regimes are identified as i. subthreshold range, ii. curve fitting range, and iii. suprasaturation range with the threshold and saturation voltages as indicated.

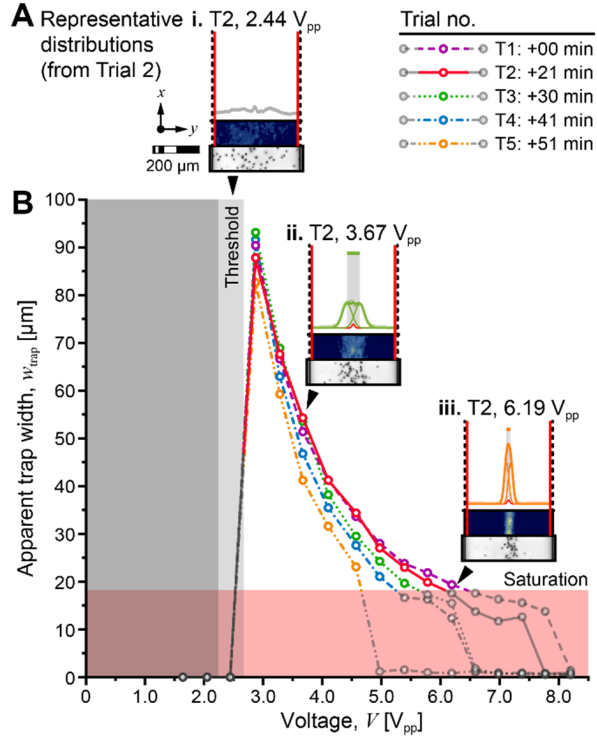

FIG. 5. (Color online). Trap width $w_{trap}$ versus applied voltage $V$. As the applied voltage increases, the field strength increases, and the trap width decreases exponentially. Representative density distributions are shown for i. threshold, ii. curve fitting range, and iii. saturation voltages for experimental Trial 2 (starting 21 min after sample preparation).

### 4. *Measurement of the acoustic energy density using polystyrene beads*

The acoustic energy density in the same silicon-glass microchannel is measured using a conventional particle tracking approach for validation of the motile cell-based method. When the device is actuated at the first half-wavelength resonance of the channel, particles undergo an acoustophoretic motion and are moved to terminal positions at the channel midline. The particle movements are recorded at five different voltages ranging from 1.68 $V_{pp}$ to 5 $V_{pp}$. The particle velocity is extracted from the imaging data, and the balance of acoustic radiation force and drag is used to determine the acoustic energy density.

Fig. 6B shows individual particle trajectories traced using GDPT[30] at the 4.25 $V_{pp}$ voltage step. The region of interest (equivalent to the field of view) is divided into five segments, and the corresponding velocity plot of every particle enables determination of an average velocity vs. *y*-position for each segment (Fig. 6A). As before, segments are aligned to create the sinusoidal particle velocity plots for all five voltages investigated (Fig. 6C). The acoustic energy density is determined as a fitting parameter by fitting the transversal velocity distribution to the theoretical prediction obtained by inserting the transverse radiation force $F_y^{rad}$ in Eq. (6) into the non-swimming and transverse version of the particle motion equation [Eq. (2)],

$$u_y = \frac{2\Phi}{3\eta} a^2 k_y E^{ac} \sin(2k_y y). \qquad (9)$$

The resultant energy density as a function of applied voltage is shown in Fig. 6D. The measured acoustic energy density very closely fits the expected power law $E^{ac} \propto V^2$, confirming that the energy density is proportional to the applied voltage squared. Note that the microfluidic channel walls impose a hydrodynamic drag on the polystyrene beads undergoing an acoustophoretic movement. The wall-drag correction factor, a function of the channel height, particle location in the $z$ direction, and particle size, was incorporated.[35, 36] The final relationship incorporating the wall-drag correction is $E^{ac} = 2.09\ V^2$.

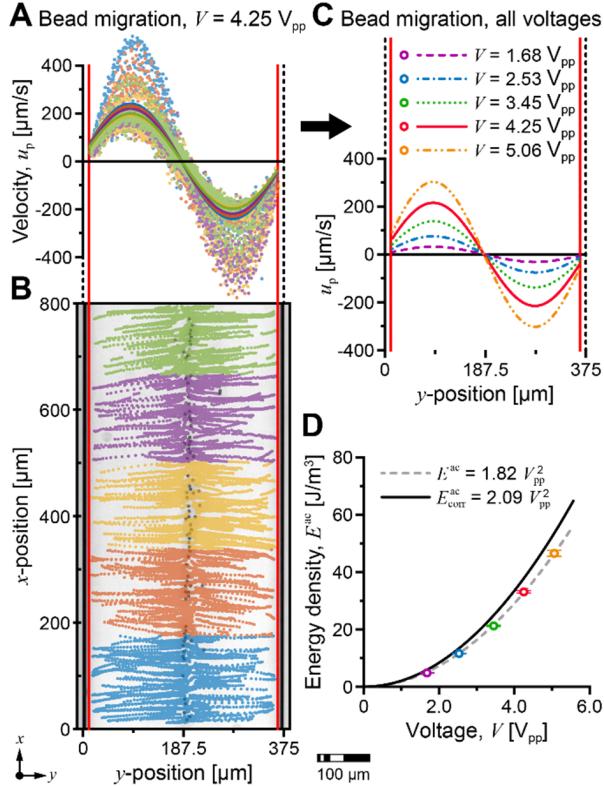

FIG. 6. (Color online). Acoustic field measurement using passive polystyrene particles. (A) Particle velocity for every particle across the channel width. (B) Particle trajectories at the first half-wavelength resonance of the rectangular microchannel. (C) Average particle velocity at five different voltages: 1.68 $V_{pp}$, 2.53 $V_{pp}$, 3.45 $V_{pp}$, 4.25 $V_{pp}$, and 5.06 $V_{pp}$. (D) Acoustic energy density $E^{ac}$ as a function of the applied voltage $V$.

### 5. Motile cell-based method validation

The trends in $E^{ac}$ vs. $V$ obtained using the two different methods are almost identical (Fig. 7). Due to the potential variation in the swim speed (and consequently the swim force) over time, the motile cell-based experimental trials are divided into three groups based on the time elapsed from the start of Trial 1. The first trial exhibits exactly the same $E^{ac}$ vs. $V$ relationship obtained from the passive particle tracing, with the leading coefficient of the former deviating by less than 1 % from the reference value (Fig. 7A). The second trial, performed approximately 20 min after the first, also provides a good match to results from the conventional bead-based method (Fig. 7B). This is a remarkable outcome given the typical inconsistencies of biological cells (as evidenced by variations in size, material properties, and here, swimming characteristics). Further, *C. reinhardtii* cells have the potential to become a rapid and accurate experimental measurement tool in acoustofluidics, replacing laborious and time-consuming approaches involving passive particles. Trials 3–5, performed more than 30 minutes after the first trial, show a slight deviation from the reference result (and those of Trials 1 and 2) (Fig. 7C). It is unclear if this is due to changes in swimming behavior or operating condition, but there is a clear drift toward an overprediction of the $E^{ac}$. Regardless the source of this change in behavior, the cell-based method could account for any such changes by taking *in situ* swimming speed measurements prior to each experimental trial. Wild-type *C. reinhardtii* allow measurement of acoustic energy densities in the 0 to 100 J m$^{-3}$ range, which is relevant to many acoustic microfluidic devices. The pressure amplitude can be simply derived from the definition of acoustic energy density $E^{ac} = p_a^2/(4\rho_o c_o^2)$ and is in the range from 0 to ~1 MPa. Again, this demonstrates that *C. reinhardtii* can be broadly applicable in the field of acoustofluidics.

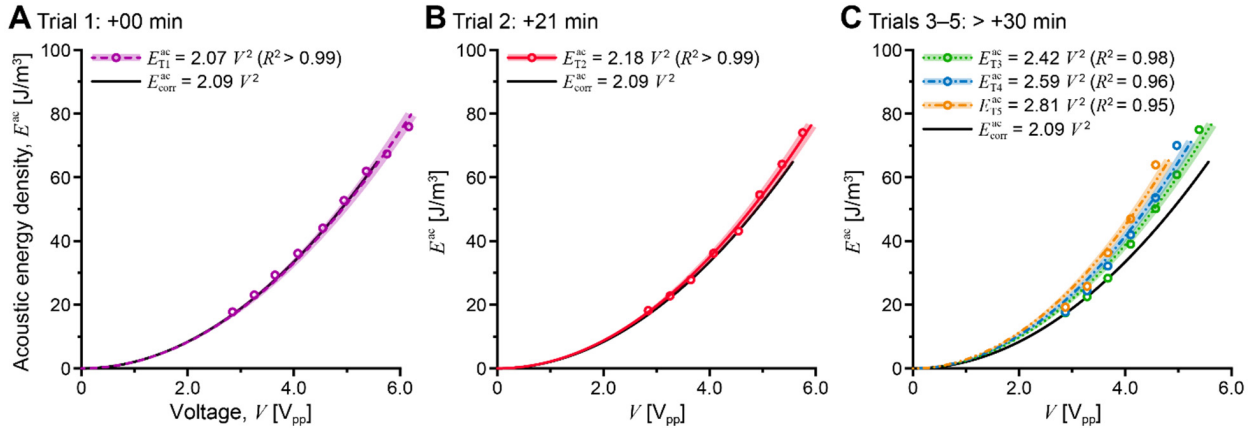

FIG. 7. (Color online). Validation of the motile cell-based performance characterization method by comparing the $E^{ac}$ vs. $V$ relationships obtained using *C. reinhardtii* measurement probes and passive particle tracing. Motile cell-based measurement (A) Trial 1: +00 min, (B) Trial 2: +21 min, and (C) Trials 3–5: > +30 min.

## IV. CONCLUSION

In this work, we demonstrate that *Chlamydomonas reinhardtii* cells can be used as active probes to accurately measure the acoustic energy density in acoustofluidic devices. The acoustophysical properties of *C. reinhardtii* cells and their growth medium needed for the field quantification are also reported. Based on the experimentally determined properties, the acoustic energy density at different field strengths is calculated in a straight channel at the first half-wavelength resonance. The resulting relationship between acoustic energy density and drive voltage closely matches (within 1 %) that measured under the same conditions using a conventional (and less-efficient) method that relies on passive polymer beads. The use of *C. reinhardtii* cells instead of passive beads significantly shortens the field characterization and quantitative analysis time, from several hours to several minutes. Not only does this reduce the length of experiments, but it also provides an improved and necessary understanding of parametric sensitivities to operating conditions that occur at different time scales. Since *C. reinhardtii* cells are algae cells, they are inexpensive, accessible, and easy to maintain. All these properties suggest the potential to develop *C. reinhardtii* as a standardized, active probe system to measure device performance. By establishing this method, we hope to improve the operational stability of acoustofluidic technologies, accelerating their adoption in emerging application areas. We are currently exploring method applicability to more complex acoustic wave fields (including propagating waves). In the future, the method may be extended to performance characterization of other microfluidic manipulation/separation technologies (e.g., electrophoretic, magnetophoretic, and optical).

**ACKNOWLEGEMENTS**

This work was supported by the National Science Foundation under Grant Nos. CMMI-1633971 and CBET-1944063. MK was partially supported by the Spencer T. and Ann W. Olin Fellowship. The authors acknowledge partial financial support from Washington University in St. Louis and the Institute of Materials Science and Engineering for the use of fabrication instruments and staff assistance. The authors would also like to thank the Dutcher lab for providing *C. reinhardtii* cells and sharing their expertise on the cells.

36. J. Happel, and H. Brenner, *Low Reynolds number hydrodynamics: with special applications to particulate media* (Springer Science & Business Media, 2012), vol. 1.18